\documentclass[twocolumn,aps,prb,superscriptaddress,a4paper,nofootinbib,showpacs,showkeys,lengthcheck]{revtex4-1}

\setcitestyle{numbers,square}

\usepackage{graphicx}
\usepackage{type1cm}
\usepackage[normalem]{ulem}  

\usepackage{bm,dcolumn,amssymb,amsmath,braket,siunitx,color}
\usepackage{sidecap}

\newcommand{\mm}[1]{\mbox{$#1$}}
\newcommand{\dstd}{\mathrm{d}}

\newcommand{\aex}{\mbox{$A_{\text{ex}}$}}

\newcommand{\ea}{{\it et al.}}

\begin{document}

\title{Spin wave stiffness and exchange stiffness of doped permalloy
  via ab-initio calculations}



\author{O. \surname{\v{S}ipr}} 
\email{sipr@fzu.cz}
\homepage{http://www.fzu.cz/~sipr} \affiliation{ Institute of Physics
  of the Czech Academy of Sciences, Cukrovarnick\'{a}~10,
  CZ-162~53~Prague, Czech Republic }  \affiliation{New Technologies Research
  Centre, University of West Bohemia, Pilsen, Czech Republic}

\author{S. \surname{Mankovsky}} \affiliation{Universit\"{a}t
  M\"{u}nchen, Department Chemie, Butenandtstr.~5-13,
  D-81377~M\"{u}nchen, Germany}

\author{H. \surname{Ebert}} \affiliation{Universit\"{a}t M\"{u}nchen,
  Department Chemie, Butenandtstr.~5-13, D-81377~M\"{u}nchen, Germany}

\date{\today}

\begin{abstract}
The way doping affects the spin wave stiffness and the exchange
stiffness of permalloy (Py) is investigated via ab-initio
calculations, using the Korringa-Kohn-Rostoker (KKR) Green function
formalism.  By considering various types of dopants of different
nature (V, Gd, and Pt), we are able to draw general conclusions.  To
describe the trends of the stiffness with doping it is sufficient to
account for the exchange coupling between nearest neighbors.  The
polarizability of the impurities is not important for the spin wave
stiffness. Rather, the decisive factor is the hybridization between
the impurity and the host states as reflected by changes in the Bloch
spectral function.  Our theoretical results agree well with earlier
experiments.
\end{abstract}



\maketitle


\section{Introduction}   \label{sec-intro}

The modification of the properties of magnetic materials by doping is
a promising and intensively studied way to make progress in device
technology and spintronics.  One of the widely studied materials in
this respect is permalloy Fe$_{19}$Ni$_{81}$.  It attracts attention
because of its high magnetic permeability but also because of its
transport properties, which are characterized by a high and low
electrical conductivity in the majority and minority spin channels,
respectively.

When studying the spin dynamics of materials, the continuum
approximation can be often employed.  It assumes that the angle
of the magnetization changes slowly over atomic distances so that the
spin vectors can be replaced by a continuous
function~\mm{\bm{m}(\bm{r})}.  The exchange energy is then
\cite{VBL+08}
\begin{equation}
E[\bm{m}] \: = \: \int_{V} \!
\dstd^{3} \bm{r} \, A_{\text{ex}} 
\sum_{c=x,y,z}
\left(\frac{\partial \bm{m}}{\partial c}\right)^{2}
\quad .
\end{equation}
Here, the exchange stiffness constant $A_{\text{ex}}$ is closely
connected to the spin wave stiffness constant $D$ via \cite{VBL+08}
\begin{equation}
  A_{\text{ex}} \; = \; \frac{ D \, M_{s}}{ 2 g \mu_{B}}
  \quad ,
\label{eq-A-from-D}
\end{equation}
where $M_{s}$ is the saturation magnetization, $g$ is the Land\'{e}
factor ($g \approx 2$ for metals) and $\mu_{B}$ is the Bohr magneton.
The spin wave stiffness constant $D$ is linked to the long wavelength
limit of the acoustic mode of magnon dispersion \cite{Kub+00},
\begin{equation}
\epsilon(\bm{q}) \; = \;
  D \, |\bm{q}|^{2} \: + \:
  \beta \, |\bm{q}|^{4} \: + \: \ldots
\quad ,
\end{equation}
where $\epsilon(\bm{q})$ is the magnon energy and $\bm{q}$ is the
corresponding wave vector.
 From the application point of view, the exchange stiffness is an
  important characteristics as it determines --- together with the
  magnetic anisotropy --- the domain structure of magnetic materials. 
  In addition, it determines magnetization dynamics in them.

Recent interest in the spin wave and/or exchange stiffness of doped
permalloy (Py) has been motived both by fundamental science and by
potential applications.  Yin \ea\ \cite{YPA+15} studied the magnetic
properties of Py doped with noble metals Ag, Pt, and Au and found a
good agreement between theoretical and experimental values of
\aex\ for dopant concentrations 10--30~\%.  In another study the
effect of doping Py by 4$d$ and 5$d$ elements on the stiffness was
studied theoretically for modest concentrations (5--15~\%) but
corresponding experimental data for comparison are lacking
\cite{PCH+16}.  On the other hand, experimental data are available for
Py doped with V, Gd, and Pt with concentrations from 1~\% to 10~\%
thanks to the works of Lepadatu \ea\ \cite{LCK+10,LCC+10} and Hrabec
\ea\ \cite{HGS+16}.

Tailoring magnetic properties by alloying and doping is one of the
very active fields in materials science.  It would be useful to
understand better the mechanism how different dopants affect the spin
wave and exchange stiffness.  In particular, it was suggested that the
polarizability of the dopant may be an important factor
\cite{YPA+15,HGS+16}. This should be checked and assessed against
other possible factors, such as hybridization between host and dopant
states.  Another question is whether the trends of the spin wave
stiffness $D$ and the exchange stiffness \aex\ with the composition of
an alloy system are always the same.  The studies often focus either
on one quantity or the other.  Given the relation
Eq.~(\ref{eq-A-from-D}) between $D$ and \aex\, this may be appropriate
but then again it is possible that some dopants will have the same
influence on $D$ but different influence on $M_{s}$, implying that the
trends of $D$ will be similar for these dopants but the trends of
\aex\ will differ.  Finally, for practical applications it would be
helpful to have a simple model or {\em ansatz} that would enable a
quick estimate how the stiffness will change upon a particular doping.

To address these questions, we present here results of an {\it
  ab-initio} study of the spin wave stiffness and the corresponding
exchange stiffness for Py doped with V, Gd, and Pt.  These dopants
have quite different electronic and magnetic properties in many
respects.  Here, the electronic structure is calculated relying on
spin-density functional theory.  For the Gd dopant the open core
formalism is employed.  The exchange coupling constants are then
obtained and the $D$ and \aex\ constants are evaluated and compared to
experiment. The influence of the magnetic properties of the dopants is
discussed and analyzed.  Moreover, it is shown that the effect of
doping on $D$ can be parametrized either in terms of the mean-field
critical temperature or in terms of the nearest-neighbor coupling.
Densities of states and Bloch spectral functions are evaluated and
compared to monitor the hybridization between the host and dopant
states.


\section{Theoretical scheme}    \label{sec-method}


\subsection{Evaluation of the stiffness constants}

\label{sec-equations}

 We assume that 
 the system can be described by a Heisenberg Hamiltonian
\begin{equation}
 H \: = \: - \, \sum_{ij} J_{ij} \,
 \bm{\hat{e}}_{i} \cdot  \bm{\hat{e}}_{j}
 \quad ,
 \label{eq-heisen}
\end{equation}
with $\bm{\hat{e}}_{i}$ being a unit vector characterizing the
orientation of the magnetic moment for the atom $i$ and $J_{ij}$ being
the exchange coupling constant.  The spin wave stiffness constant $D$
can then be expressed as a sum \cite{LKAG87,PKT+01}
\begin{equation}
  D \: =\: \sum_{j} 
  \frac{2 \mu_{B}}{3 \mu_{j}} \,
  J_{0j} \, R_{0j}^{2}
\quad ,
  \label{eq-simple}
\end{equation}
where $\mu_{j}$ is the magnetic moment of atom $j$ and $R_{0j}$ is the
corresponding inter-atomic distance.  Eq.~(\ref{eq-simple}) was
originally derived for systems containing atoms of one type only.  The
sum over the atoms $\sum_{j}$ converges only conditionally, so an
additional damping factor has been introduced which enables evaluation
of Eq.~(\ref{eq-simple}) by extrapolating the partial results to zero
damping \cite{PKT+01}.  Recently the applicability of
Eq.~(\ref{eq-simple}) was extended to multicomponent systems
\cite{TCF+09,DGC+15,PCH+16}.  In our case we are dealing with doped Py
so we have atoms of three different types located on lattice sites of
an fcc structure.  If we label atomic types by $\alpha$ and lattice
sites by $j$, the spin wave stiffness constant $D$ can be evaluated as
\begin{eqnarray}
  D & = &  \lim_{\eta \rightarrow 0} D(\eta)
  \; ,  \label{eq-dlim} \\
  D(\eta) & = & \sum_{\alpha} c_{\alpha} \, D_{\alpha}(\eta)
  \; ,  \label{eq-dalpha} \\
  D_{\alpha}(\eta) & = &  \sum_{j} \sum_{\beta} \,
  c_{\beta} \, 
  \frac{2 \mu_{B}}{3\sqrt{|\mu_{\alpha}| |\mu_{\beta}|}} \,
  J_{0j}^{(\alpha\beta)} \, R_{0j}^{2} \,
  \mathrm{e}^{-\eta \frac{R_{0j}}{R_{01}}}
  \; ,
   \label{eq-dsum}
\end{eqnarray}
where $c_{\alpha}$ and $\mu_{\alpha}$ are the concentration and the
magnetic moment of atoms of type $\alpha$, $J_{0j}^{(\alpha\beta)}$ is
the pairwise exchange coupling constant if an atom of type $\alpha$ is
located at the lattice origin and an atom of type $\beta$ is
located at the lattice site $j$, $R_{0j}$ is the distance of the site
$j$ from the lattice origin, $\eta$ is the damping parameter and $R_{01}$
is the nearest-neighbor interatomic distance.

When evaluating the sums in Eqs.~(\ref{eq-dalpha})--(\ref{eq-dsum}),
one should keep in mind that the magnetic moments are not all of the
same nature.  Namely, the moments of the V and Pt atoms do not
originate from the atoms themselves but are induced by the neighboring
Fe and Ni atoms.  Thus, they cannot be treated as independent entities
when describing the tilting of the magnetic moments.  This point was
recognized in the past, for example, in connection with efforts to
describe the temperature-dependence of magnetism of compounds such as
FePt \cite{MNG+05}, FeRh \cite{Mry+05} or NiMnSb \cite{LME+06}.  A
thorough discussion and a way to solve the problem by treating the
induced moments via the linear response formalism can be found in
\cite{PMS+10}.  In our case the situation is different because we are
interested in the spin wave stiffness for $T$=0, which is determined
by the energetics of spin waves in the long-wave-length limit, where
the angles between the spins are very small and the corresponding
decrease of the induced magnetic moments will be also very small.

Reckoning all this, we deal with the moments on V and Pt atoms in a
hybrid way: we include them in the $\sum_{\beta}$-sum in
Eq.~(\ref{eq-dsum}) but not in the $\sum_{\alpha}$-sum in
Eq.~(\ref{eq-dalpha}).  The V and Pt moments are thus supporting the
orientations of magnetic moments at the Fe and Ni atoms but they
themselves do not contribute to $D$ directly.  To get more insight
into the role of the dopant moments, we present further on in
Sec.~\ref{sec-the-exp} also the results obtained when the moments at
the dopants were ignored completely in
Eqs.~(\ref{eq-dalpha})--(\ref{eq-dsum}) and when they were treated
equally as the moments at Fe or Ni atoms, i.e., fully included both in
the $\sum_{\alpha}$-sum in Eq.~(\ref{eq-dalpha}) and in the
$\sum_{\beta}$-sum in Eq.~(\ref{eq-dsum}). The moments of Gd atoms are
intrinsic.  Therefore, we treat them in the same way as moments of Fe
and Ni atoms, unless explicitly said otherwise.

The exchange stiffness constant $A_{\text{ex}}$ was obtained from the
spin wave stiffness constant $D$ using Eq.~(\ref{eq-A-from-D}).  The
saturation magnetization $M_{s}$ was determined as the magnetic moment
per unit cell (always including the contributions of all atomic
types).  The Land\'{e} factor was taken as $g=2.1$ \cite{NLC+03,SNS+13}.


\subsection{Computational method}    \label{sec-comput}

The calculations were performed within the ab-initio framework of the
spin-density functional theory, relying on the generalized gradient
approximation (GGA) using the Perdew, Burke and Ernzerhof (PBE)
functional. The electronic structure was calculated in a
scalar-relativistic mode using the spin-polarized multiple-scattering
or Korringa-Kohn-Rostoker (KKR) Green function formalism \cite{EKM11}
as implemented in the {\sc sprkkr} code \cite{sprkkr-code}.
For the multipole expansion of the Green function, an
angular momentum cutoff \mm{\ell_{\mathrm{max}}}=3 was used.
The
disorder was treated within the coherent potential approximation
(CPA).  The potentials were subject to the atomic sphere approximation
(ASA).  
 Identical atomic radii were used for all atomic species on a
  given site, as it is common in CPA calculations.  By doing this we
  neglect effects of local lattice relaxations and may effectively
  introduce some artificial charge transfer \cite{KD+90}.  We do not
  expect that this affects our conclusions significantly. In
  principle, these constraints could be by-passed by employing unequal
  constituent atoms radii \cite{CKM+16}.
For each
dopant concentration, the equilibrium lattice constant $a_{0}$ was
determined by minimizing the total energy.  The exchange coupling
constants $J_{0j}^{(\alpha\beta)}$ were evaluated from the electronic
structure using the prescription of Liechtenstein \ea\ \cite{LKAG87}.

Taking the limit $\lim_{\eta \rightarrow 0} D(\eta)$ in
Eq.~(\ref{eq-dlim}) is a delicate issue.  To avoid errors in
extrapolating $D(\eta)$ to $\eta=0$, one should evaluate $D(\eta)$
down to as low $\eta$ as possible.  However, low $\eta$ implies that
the sum Eq.~(\ref{eq-dsum}) converges slowly with the distance
$R_{0j}$.  One should, therefore, extend the sum $\sum_{j}$ to large
distances.  Evaluating the exchange coupling constants
$J_{0j}^{(\alpha\beta)}$ for large $R_{0j}$ requires a very dense mesh
in the $\bm{k}$-space to avoid numerical errors for the structure
constants.  In our case the situation is not so critical because we
are dealing with alloys, meaning that the requirements on the extent
of the $\sum_{j}$-sum in Eq.~(\ref{eq-dsum}) and on the density of the
$\bm{k}$-mesh in evaluating the $J_{0j}^{(\alpha\beta)}$ constants are
not so demanding.  Nevertheless, to be on the safe side, the
$\bm{k}$-space integration was carried out via sampling on a regular
mesh corresponding to 62$\times$62$\times$62 points in the full
Brillouin zone and the $\sum_{j}$-sum in Eq.~(\ref{eq-dsum}) covered
interatomic distances up to 20.5~$a_{0}$ (corresponding to about
136000 atomic sites).  With these settings numerically accurate values
of $D(\eta)$ were obtained for $\eta$ ranging from 0.2 to 1.  The
extrapolation to $\eta=0$ was done using a fifth-degree polynomial.

A proper description of magnetism of Gd requires going beyond the
GGA. Many aspects of magnetism of rare earth metals can be described
within the open-core formalism, where the $f$ electrons are treated as
tightly bound core electrons and their number is kept fixed to an
integer number during the self-consistency loop \cite{Rich+01}.  In
particular it was shown that the exchange coupling of Gd and its
compounds can be described by the open core formalism quite well
\cite{TKB+03,RTD+05,KKR+07}. Therefore we employ it here when
discussing the impact of doping by Gd atoms.  More specifically, we
used a mixed approach where we first calculate the electronic
structure of Gd-doped Py via the open core formalism and then we use
the self-consistent potential obtained thereby to evaluate the
exchange coupling constants in a standard way, i.e., treating the Gd
$f$-electrons as valence electrons. To check whether this approach is
justified, we compared the density of states (DOS) obtained via both
approaches.  We found that the positions of the Gd $f$-states
practically do not depend on whether they are provided by the open
core calculation itself or whether they are derived from the peaks in
the density of the Gd $f$-states obtained by a ``standard''
calculation for the potential generated by the open-core formalism
(data not shown).  It turns out in the end that use of the open core
formalism is not crucial, the stiffness constants obtained in this way
are very close to the constants obtained by relying solely on the
band-like description of the $f$-electrons.  The equilibrium lattice
constant $a_{0}$ was evaluated always within the GGA, for all dopant
types.


\section{Results}   \label{sec-res}


\subsection{Comparing calculated values for $D$ and $A_{\text{ex}}$ with experiment}  

\label{sec-the-exp}

\begin{table}
\caption{Equilibrium lattice constant $a_{0}$, magnetic moment per
  unit cell $M_{s}$, and spin waves stiffness constant $D$ for V-doped
  Py.}
\label{tab-stiff-V}
\begin{ruledtabular}
\begin{tabular}{dccc}
  \multicolumn{1}{c}{conc.} &
  \multicolumn{1}{c}{$a_{0}$} &
  \multicolumn{1}{c}{$M_{s}$} &
  \multicolumn{1}{c}{$D$}  \\
  \multicolumn{1}{c}{(\%)} & 
  \multicolumn{1}{c}{(a.u.)} &
  \multicolumn{1}{c}{($\mu_{B}$/cell)} &
  \multicolumn{1}{c}{(meV~\AA$^{2}$)} \\
\hline
 0.0  &  6.658  &  1.017   &  576 \\
 1.0  &  6.662  &  0.982   &  534 \\   
 2.5  &  6.666  &  0.926   &  478 \\  
 3.5  &  6.670  &  0.888   &  444 \\  
 6.0  &  6.676  &  0.799   &  369 \\   
 10.0 &  6.695  &  0.671   &  273 
\end{tabular}
\end{ruledtabular}
\end{table}

Results obtained for the equilibrium lattice constant $a_{0}$,
magnetization per unit cell $M_{s}$, and spin wave stiffness $D$ for
V-doped Py are shown in Tab.~\ref{tab-stiff-V}.  As discussed in
Sec.~\ref{sec-equations}, we show $D$ obtained when treating the
moments of the V atoms as supporting, i.e., omitting them in the
$\sum_{\alpha}$-sum in Eq.~(\ref{eq-dalpha}) but keeping them in the
$\sum_{\beta}$-sum in Eq.~(\ref{eq-dsum}). The magnetic moments of the
V atoms are oriented antiparallel to the moments at the Fe and Ni
atoms, therefore the magnetization $M_{s}$ decreases rapidly with
increasing concentration of the V atoms.

\begin{table}
\caption{Equilibrium lattice constant $a_{0}$, magnetic moment per unit
  cell $M_{s}$, and spin waves stiffness constant $D$ for Gd-doped Py.  The
  unit for $D$ is meV~\AA$^{2}$.}
\label{tab-stiff-Gd}
\begin{ruledtabular}
\begin{tabular}{dccccc}
 & & & \multicolumn{3}{c}{stiffness constant $D$} \\
  \multicolumn{1}{c}{conc.} &
  \multicolumn{1}{c}{$a_{0}$} &
  \multicolumn{1}{c}{$M_{s}$} &
  \multicolumn{1}{c}{dopants} &
  \multicolumn{1}{c}{dopants} &
  \multicolumn{1}{c}{dopants}  \\
  \multicolumn{1}{c}{(\%)} & 
  \multicolumn{1}{c}{(a.u.)} &
  \multicolumn{1}{c}{($\mu_{B}$/cell)} &
  \multicolumn{1}{c}{ignored} 
  &  \multicolumn{1}{c}{supporting} 
  &  \multicolumn{1}{c}{as equal} \\
\hline
  0.0  &  6.658  &  1.017  &  576  &  576  &  576 \\
  1.0  &  6.693  &  1.067  &  531  &  531  &  530 \\   
  5.0  &  6.803  &  1.244  &  380  &  381  &  381 \\  
 10.0  &  6.952  &  1.477  &  242  &  243  &  243 
\end{tabular}
\end{ruledtabular}
\end{table}

Analogous results for Gd-doped Py are shown in
Tab.~\ref{tab-stiff-Gd}.  We present here $D$ calculated by three
different methods: (i)~ignoring the moments at Gd atoms altogether
[omitting them both in the $\sum_{\alpha}$-sum and in the
  $\sum_{\beta}$-sum in Eqs.~(\ref{eq-dalpha})--(\ref{eq-dsum})],
(ii)~treating them as supporting [omitting them in the
  $\sum_{\alpha}$-sum in Eq.~(\ref{eq-dalpha}) but keeping them in the
  the $\sum_{\beta}$-sum in Eq.~(\ref{eq-dsum})], and (iii)~treating
the Gd moments equally as the Fe and Ni moments (including them in the
$\sum_{\alpha}$-sum and in the $\sum_{\beta}$-sum).  It is obvious
from Tab.~\ref{tab-stiff-Gd} that the way the Gd moments are treated
does not really matter for the spin wave stiffness of Gd-doped Py.

\begin{table}
\caption{Magnetic moments per unit cell $M_{s}$ and spin waves
  stiffness constant $D$ for Gd-doped Py obtained by taking the
  potential from the open-core calculations and from the GGA-based
  band structure calculations.}
\label{tab-gga-oc}
\begin{ruledtabular}
\begin{tabular}{dcccc}
 & \multicolumn{2}{c}{open-core potential} 
 & \multicolumn{2}{c}{GGA potential} \\
  \multicolumn{1}{c}{conc.} &
  \multicolumn{1}{c}{$M_{s}$} & \multicolumn{1}{c}{$D$} &
  \multicolumn{1}{c}{$M_{s}$} & \multicolumn{1}{c}{$D$} \\
  \multicolumn{1}{c}{(\%)} & 
  \multicolumn{1}{c}{($\mu_{B}$/cell)} &
  \multicolumn{1}{c}{(meV~\AA$^{2}$)} &
  \multicolumn{1}{c}{($\mu_{B}$/cell)} &
  \multicolumn{1}{c}{(meV~\AA$^{2}$)} \\
\hline
  1.0  &  1.067 &  531  &  1.066 &  532  \\   
  5.0  &  1.244 &  381  &  1.239 &  387  \\  
 10.0  &  1.477 &  243  &  1.463 &  250   
\end{tabular}
\end{ruledtabular}
\end{table}

The influence of using the open core formalism for Gd-doped Py is
shown in Tab.~\ref{tab-gga-oc}.  It contains the magnetization $M_{s}$
and the spin wave stiffness $D$ calculated for the potential obtained
using the open core formalism and for the potential obtained via the
GGA.  One can see that the difference between both procedures is not
significant in this regard.

\begin{table}
\caption{Equilibrium lattice constant $a_{0}$, magnetic moment per
  unit cell $M_{s}$, and spin waves stiffness constant $D$ calculated
  for Pt-doped Py.  Experimental values for $D$ of Yin
  \ea\ \cite{YAD+17} are shown in the last column.}
\label{tab-stiff-Pt}
\begin{ruledtabular}
\begin{tabular}{dcccc}
  \multicolumn{1}{c}{conc.} &
  \multicolumn{1}{c}{$a_{0}$} &
  \multicolumn{1}{c}{$M_{s}$} &
  \multicolumn{1}{c}{$D$ calc.} &
  \multicolumn{1}{c}{$D$ exper.} \\
  \multicolumn{1}{c}{(\%)} & 
  \multicolumn{1}{c}{(a.u.)} &
  \multicolumn{1}{c}{($\mu_{B}$/cell)} &
  \multicolumn{1}{c}{(meV~\AA$^{2}$)} &
  \multicolumn{1}{c}{(meV~\AA$^{2}$)} \\
\hline
  0.0  &  6.658  &  1.017  &  576  & 442  \\
  2.5  &  6.708  &  1.015  &  564  &   \\   
  5.0  &  6.755  &  1.014  &  549  &   \\  
  7.5  &  6.798  &  1.013  &  533  &   \\  
 10.0  &  6.837  &  1.007  &  519  &   \\   
 13.0  &  6.876  &  0.997  &  504  &   \\  
 15.0  &  6.905  &  0.992  &  493  & 388  \\  
 20.0  &  6.972  &  0.972  &  469  &   \\  
 30.0  &  7.078  &  0.923  &  420  & 329 
\end{tabular}
\end{ruledtabular}
\end{table}

Results for Pt-doped Py are shown in Tab.~\ref{tab-stiff-Pt}.  Here we
show additionally the experimental results for $D$ at zero temperature
obtained by Yin \ea\ \cite{YAD+17}.  The experimentally observed
decrease of $D$ with increasing Pt concentration is similar to the
decrease obtained by theory.

\begin{figure}
\includegraphics[viewport=0.5cm 0.5cm 9.0cm 15.5cm]{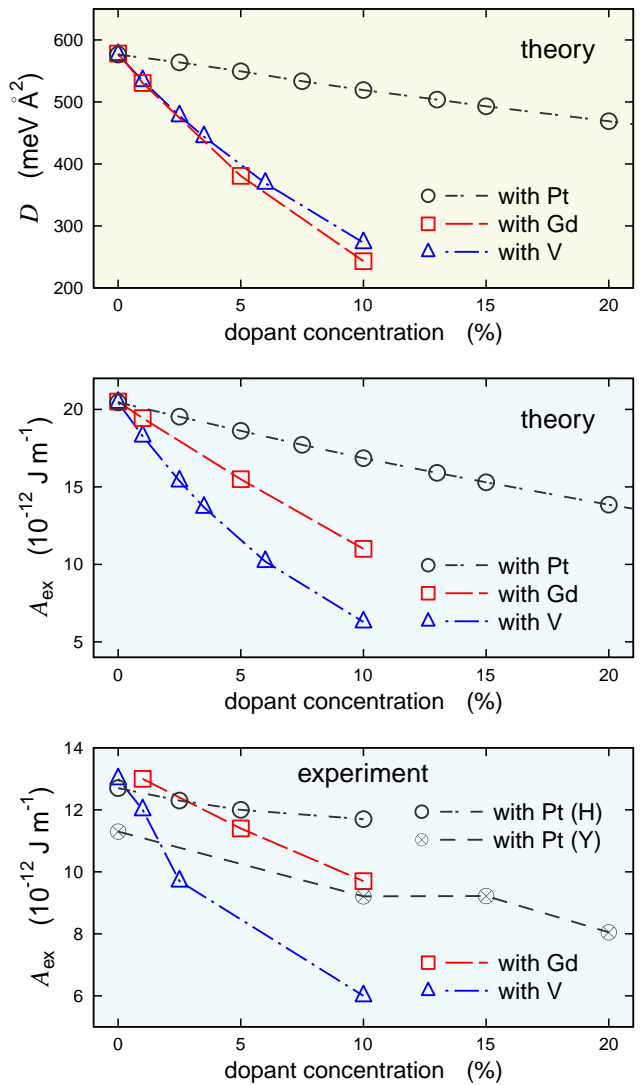}%
\caption{(Color online) Theoretical spin wave stiffness $D$ (upper
  panel) and exchange stiffness $A_{\text{ex}}$ (central panel) for Py
  doped with V, Gd, and Pt.  Experimental values for $A_{\text{ex}}$
  are shown in the lowermost panel, with the data from
  Ref.~\onlinecite{LCK+10} for V-doped Py, from
  Ref.~\onlinecite{LCC+10} for Gd-doped Py, and from
  Ref.~\onlinecite{HGS+16} (circles) and Ref.~\onlinecite{YPA+15}
  (circles with crosses) for Pt-doped Py.}
\label{fig-theory-exper}
\end{figure}

Our focus is on comparing the trends of $D$ and $A_{\text{ex}}$ with
the dopant concentration for different dopant types.  This is
inspected in Fig.~\ref{fig-theory-exper}.  One can see that the spin
wave stiffness $D$ for V-doped Py and Gd-doped Py is practically the
same.  The difference appears only for the exchange stiffness
$A_{\text{ex}}$ and stems from the differences in the magnetization
for these two systems (cf.~the third columns of
Tabs.~\ref{tab-stiff-V} and \ref{tab-stiff-Gd}).  The experimental
data are shown in the bottom graph of Fig.~\ref{fig-theory-exper}.
Both experiment and theory suggest that the decrease of the exchange
stiffness $A_{\text{ex}}$ with increasing dopant concentration is
approximately linear and that this decrease is quickest for the V
dopant and slowest for the Pt dopant.

Concerning the absolute values of $A_{\text{ex}}$, part of the
difference between theory and experiment is due to the temperature
effects.  The theoretical data are for zero temperature.  The
experimental data were obtained by fitting the temperature-dependence
of the magnetization to the Bloch law while assuming implicitly that
the stiffness itself does not depend on temperature
\cite{LCK+10,LCC+10,HGS+16}.  This is, however, not the case
\cite{MW+68,Rie+77,KB+94,YAD+17} meaning that the experimental values
for \aex\ in Fig.~\ref{fig-theory-exper} correspond to an unknown {\em
  effective} temperature.  Presumably the effect of this will not be
very large: Temperature-dependent measurements of the spin wave
stiffness of doped Py indicate that at room temperature the constant
$D$ decreases to about 90~\% of its zero-temperature value
\cite{YAD+17}.
 Therefore we assume that most of the difference in the absolute
  values of \aex\ as given by theory and by experiment comes from the
  spin wave stiffness constant $D$.  Further comments are given at the
  beginning of Sec.~\ref{sec-diskuse} below.

\begin{figure}
\includegraphics[viewport=0.5cm 0.5cm 9.0cm 5.5cm]{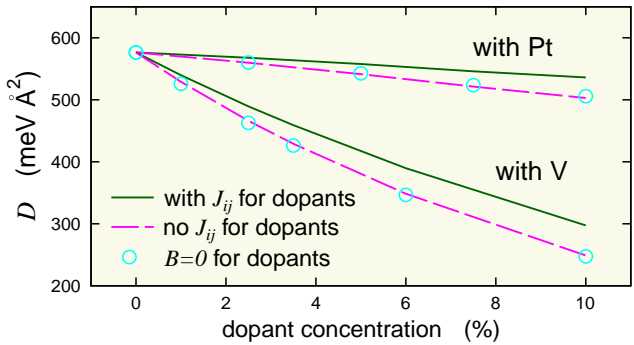}%
\caption{(Color online) Spin wave stiffness $D$ for Py doped with V
  and Pt calculated by considering the coupling constants
  $J_{ij}^{(\alpha\beta)}$ in
  Eqs.~(\protect\ref{eq-dalpha})--(\protect\ref{eq-dsum}) for all
  atoms on the same footing (full lines), by completely ignoring the
  coupling constants for the dopants (dashed lines), and by
  additionally suppressing the exchange field $B$ for the dopant atoms
  (circles).}
\label{fig-howto}
\end{figure}

One of the intensively studied fundamental questions in connection with
the spin dynamics of doped ferromagnets has been about the role of the
magnetism of dopant atoms on the exchange and spin wave stiffness
\cite{YPA+15,HGS+16}.  We have shown already that the difference in
the exchange stiffness \aex\ between V-doped and Gd-doped Py is
because of the difference in the magnetization of these systems, which
in turn stems from different magnetic moments at Gd and V atoms.  We
have also shown that for Gd doping it does not matter for $D$ whether
the Gd moments are included in Eqs.~(\ref{eq-dalpha})--(\ref{eq-dsum})
or not (Tab.~\ref{tab-stiff-Gd}).  Let us now turn to the spin wave
stiffness for Py doped with V and Pt, both of which are easy to
polarize, and compare $D$ evaluated via two extreme approaches: (i)~by
treating the dopants equally as the host atoms concerning the
coupling, i.e., both sums $\sum_{\alpha}$ in Eq.~(\ref{eq-dalpha}) and
$\sum_{\beta}$ in Eq.~(\ref{eq-dsum}) include also the dopant atoms,
and (ii)~by ignoring the coupling for the dopants altogether, i.e.,
neither of the sums $\sum_{\alpha}$ and $\sum_{\beta}$ includes the
dopants.  The difference between both approaches can be seen
Fig.~\ref{fig-howto}, where we display the results for Pt- and
V-doping only; the data for Gd-doped Py are very similar to V-doped Py
(cf.\ the upper panel of Fig.~\ref{fig-theory-exper}).  One can see
that even though ignoring the coupling for the dopants decreases the
values of $D$ by up to 20~\% (depending on the concentration), the
overall trend and especially the difference between the effect of both
dopants does not change.  To proceed further, we performed another
calculation where the effective local exchange field $B$ was
suppressed for the Pt and V dopants.  The results (depicted by the
circles in Fig.~\ref{fig-howto}) are nearly identical to the situation
when the exchange coupling involving the dopant atoms is ignored.

We conclude from this that the polarizability of the dopants does not
have a significant influence on the spin wave stiffness $D$ of doped
permalloy.  The reason why different dopants lead to different results
for $D$ must be elsewhere, presumably in the details of the electronic
structure and the related hybridization (see Sec.~\ref{sec-eledos}
below).


\subsection{Influence of dopants on the exchange coupling}  

\label{sec-jij}

\begin{figure}
\includegraphics[viewport=0.6cm 0.6cm 9.5cm 7.5cm]{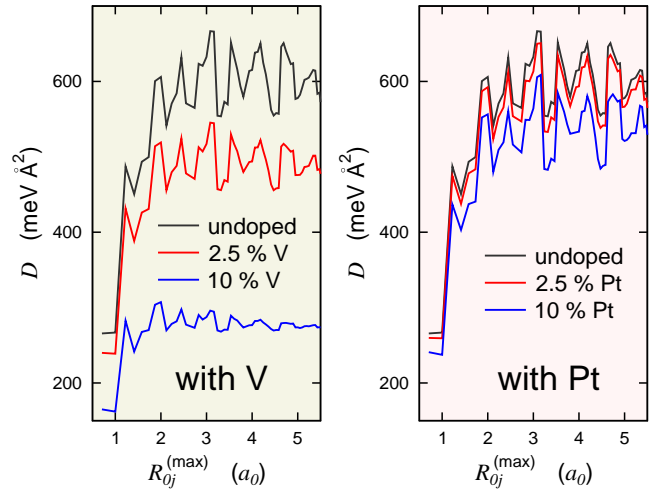}%
\caption{(Color online) Dependence of the stiffness constant $D$ on
  the maximum distance $R_{0j}^{\text{(max)}}$ accounted for when
  evaluating Eq.~(\ref{eq-dsum}) for undoped permalloy and for
  permalloy doped by 2.5~\% and 10~\% of V (left) and Pt (right).}
\label{fig-DR}
\end{figure}

In the following we consider the mechanism through which different
dopants affect the stiffness of permalloy.  The first information
concerns the size of the region which effectively determines the spin
wave stiffness constant $D$.  We mentioned already that to get good
numerical stability, the sum over the sites $\sum_{j}$ in
Eq.~(\ref{eq-dsum}) has to include sites at very large distances
$R_{0j}$ \cite{PKT+01}.  For a deeper insight, the dependence of the
stiffness constant $D$ on the maximum distance $R_{0j}^{\text{(max)}}$
up to which the sum $\sum_{j}$ in Eq.~(\ref{eq-dsum}) extends is
presented in Fig.~\ref{fig-DR}.  No damping has been considered for
simplicity ($\eta$=0).  The constant $D$ oscillates with
$R_{0j}^{\text{(max)}}$ and the amplitude of these oscillations decays
very slowly.  This is why the sum in Eq.~(\ref{eq-dsum}) has to cover
large distances $R_{0j}$ and why the damping factor $\eta$ has been
introduced \cite{PKT+01}. 

It is evident from Fig.~\ref{fig-DR} that significant variations of
the spin wave stiffness constant $D$ only occur within few nearest
shells --- up to about 2$a_{0}$ ($a_{0}$ is the lattice constant).
Afterwards, $D$ just oscillates around the mean value.  If the
difference in $D$ for two dopant concentrations is big for large
$R_{0j}^{\text{(max)}}$, it is big also for small
$R_{0j}^{\text{(max)}}$ and {\em vice versa}. The difference between V
and Pt concerning the rate how $D$ decreases with increasing dopant
concentration (see Fig.~\ref{fig-theory-exper}) is evident for small
values of $R_{0j}^{\text{(max)}}$ already. Including large distances
$R_{0j}$ when evaluating Eq.~(\ref{eq-dsum}) is thus necessary just
for technical reasons --- to ensure the numerical stability.
 
\begin{figure}
\includegraphics[viewport=0.6cm 0.6cm 9.5cm 11.0cm]{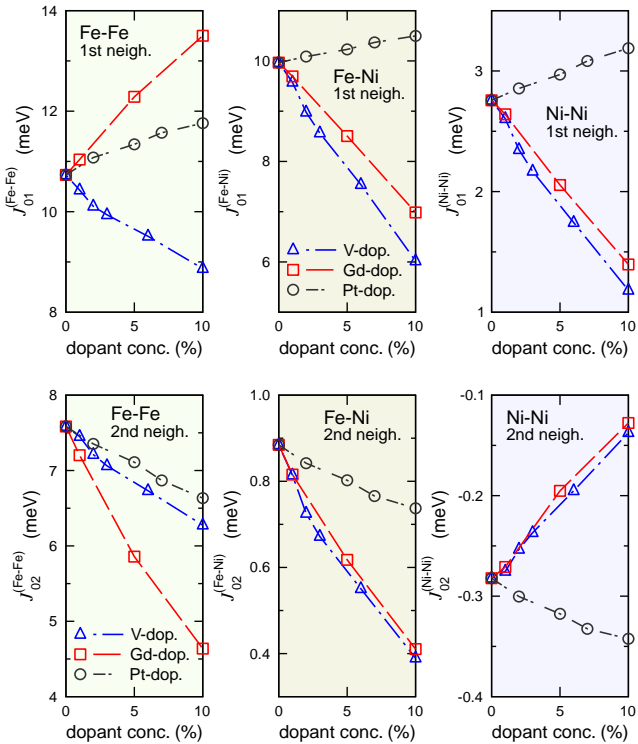}%
\caption{(Color online) Exchange coupling constants for Fe-Fe, Fe-Ni,
  and Ni-Ni atomic pairs if they are first nearest neighbors (upper
  panels) or second  nearest neighbors (lower panels). Data for
  V-doped Py are shown by triangles, for Gd-doped Py by squares, and
  for Pt-doped Py by circles.}
\label{fig-jij}
\end{figure}

Further information how the nearest neighborhood affects the spin wave
stiffness comes from the exchange coupling constants.
Fig.~\ref{fig-jij} shows them for the (Fe-Fe), (Fe-Ni) and (Ni-Ni)
pairs if they form the nearest neighbors and the next-nearest
neighbors.  Even though the stiffness constant $D$ has to be evaluated
including also pairs whose members are much further apart, the main
feature can be seen from $J_{01}$ and $J_{02}$ already: For the
(Fe-Ni) and (Ni-Ni) pairs, the coupling constants for V-doped and
Gd-doped Py are very similar whereas for Pt-doped Py they differ.  As
Ni has by far the highest concentration in our systems, this explains
why the spin wave stiffness constant $D$ is similar for the V and Gd
dopants and different for the Pt dopant.

\begin{SCfigure*}
\includegraphics[viewport=0.6cm 0.5cm 13.0cm 10.5cm]{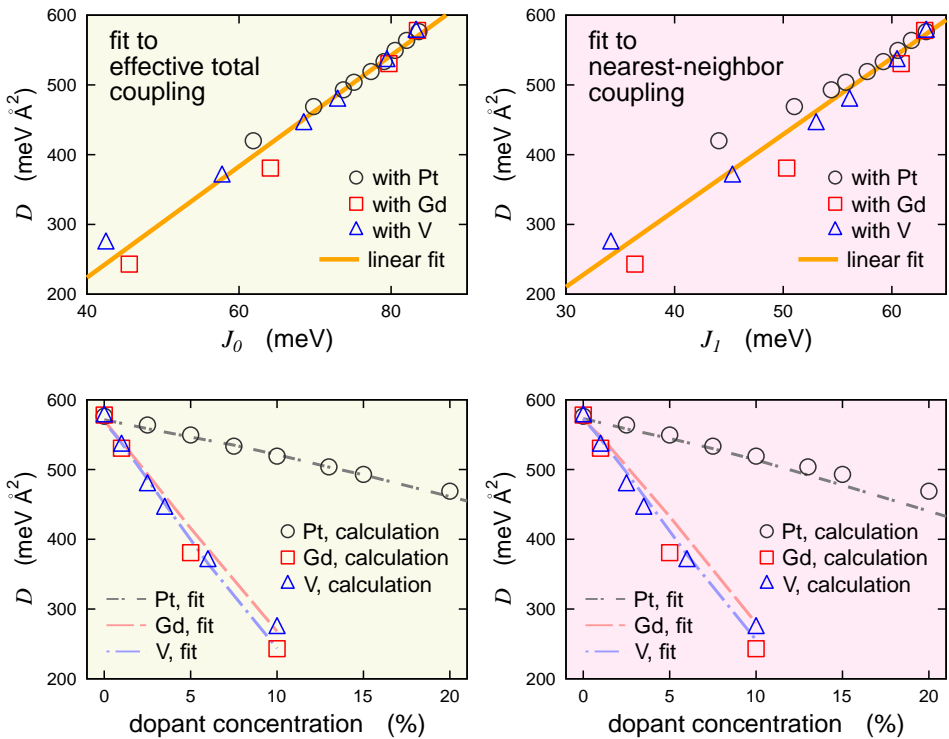}%
\caption{(Color online) Upper panels: Dependence of the theoretical
  spin wave stiffness $D$ on the effective exchange coupling parameter
  $J_{0}$ (left) and on the nearest-neighbor coupling $J_{1}$ (right).
  The markers denote calculated values of $D$, $J_{0}$, and $J_{1}$
  for different dopants and concentrations, the straight lines show
  the linear fits. Lower panels: Calculated spin wave stiffness $D$
  (markers) together with $D$ obtained from respective $J_{0}$ values
  relying on the linear $D(J_{0})$ fit (left) and from respective
  $J_{1}$ values relying on the \mm{D(J_{1})}\ fit (right).}
\label{fig-D-J-fit}
\end{SCfigure*}

Based on this, one can speculate that even though a formal evaluation
of $D$ requires performing the sum $\sum_{j}$ in Eq.~(\ref{eq-dsum})
up to a large distance $R_{0j}^{\text{(max)}}$, the {\em bare
  influence of the doping} might be accounted for by a more simple
quantity.  Let us first focus on the effective total coupling $J_{0}$
defined as
\begin{equation}
  J_{0} \: = \: \sum_{j} J_{0j}
  \quad ,
\label{eq-J0cryst}
\end{equation}
or, more specifically for our alloy system
[cf.\ Eqs.~(\ref{eq-dalpha})--(\ref{eq-dsum})],
\begin{equation}
  J_{0} \: = \: \sum_{\alpha} c_{\alpha}
  \sum_{j} \sum_{\beta}  c_{\beta} J_{0j}^{(\alpha\beta)}
  \quad .
\label{eq-J0}
\end{equation}
This quantity is related to the mean-field Curie temperature via
\begin{displaymath}
  T_{C}^{\text{(MFA)}} \: = \: \frac{2}{3} \, \frac{1}{k_{\text{B}}}
  \, J_{0}
  \quad .
\end{displaymath}
Note that for a translationally-periodic system, one does not need to
perform the $\sum_{j}$-sum in Eq.~(\ref{eq-J0}) explicitly, the
$J_{0}$ constant can be evaluated from the scattering paths operators
in a similar way as the individual $J_{ij}$ constants \cite{LKAG87}.

The spin wave stiffness constant $D$ is plotted as a function of the
effective total coupling $J_{0}$ (or, equivalently,
$T_{C}^{\text{(MFA)}}$) in the upper left panel of
Fig.~\ref{fig-D-J-fit}.  One can see that for Py doped with various
impurities of different concentrations the dependence of $D$ on
$J_{0}$ is nearly linear. Quantitatively it can be described by 
\begin{equation}
  D \: = \: 79.6 \, J_{0} \: - \: 95
\label{eq-fitJ0}
\end{equation}
if $J_{0}$ is in meV and $D$ in meV~\AA$^{2}$. Using this universal
fit, we can recover the dependence of $D$ on the concentration of the
impurities by first evaluating $J_{0}$ for each system and then
getting appropriate $D$ via Eq.~(\ref{eq-fitJ0}).  The outcome is
shown in the lower left panel of Fig.~\ref{fig-D-J-fit}. One can see
that this model correctly separates the trends for the Pt impurity on
the one side and for the V and Gd impurities on the other side and
that the slope of the dependency of $D$ on the concentration is also
reproduced quite well.

To simplify matters even more, one can ask whether the influence
of impurities could be possibly described just by focusing on the
nearest neighbors.  Therefore we inspect the dependence of the
stiffness constant $D$ on the effective coupling originating from the
nearest neighbors only,
\begin{equation}
  J_{1} \: = \: \sum_{\alpha} c_{\alpha} \, 12 \, 
  \sum_{\beta} c_{\beta} J_{01}^{(\alpha\beta)}
\quad .
  \label{eq-J1}
\end{equation}
The constant $J_{01}^{(\alpha\beta)}$ characterizes the exchange
coupling between the central atom of type $\alpha$ and an atom of type
$\beta$ in the first coordination shell and the factor 12 stands for
the number of nearest neighbor sites for the fcc lattice.  The
resulting dependence of $D$ on $J_{1}$ is shown in the upper right
panel of Fig.~\ref{fig-D-J-fit}.  The spread around the linear fit
\begin{equation}
  D \: = \: 109.4 \, J_{1} \: - \: 118
\label{eq-fitJ1}
\end{equation}
is now bigger than in the case of the $D(J_{0})$ dependence but still
quite small.  Indeed, by relying on Eq.~(\ref{eq-fitJ1}), one can get
the dependence of $D$ on the impurity concentration with a similar
accuracy as by relying on Eq.~(\ref{eq-fitJ0}) --- compare the lower
right panel of Fig.~\ref{fig-D-J-fit} with the lower left panel of the
same figure.   One could thus say that the {\em trend} of the
  spin wave stiffness with the dopant concentration is, to a decisive
  degree, determined by the coupling between the nearest neighbors.
  As concerns the {\em value} of the stiffness constant $D$, it
  depends on the interaction within few nearest shells --- see
  Fig.~\ref{fig-DR} and the related text at the beginning of
  Sec.~\ref{sec-jij}.


\subsection{Influence of dopants on the electronic structure}  

\label{sec-eledos}

\begin{SCfigure*}
\includegraphics[viewport=0.6cm 0.6cm 13.0cm
  10.0cm]{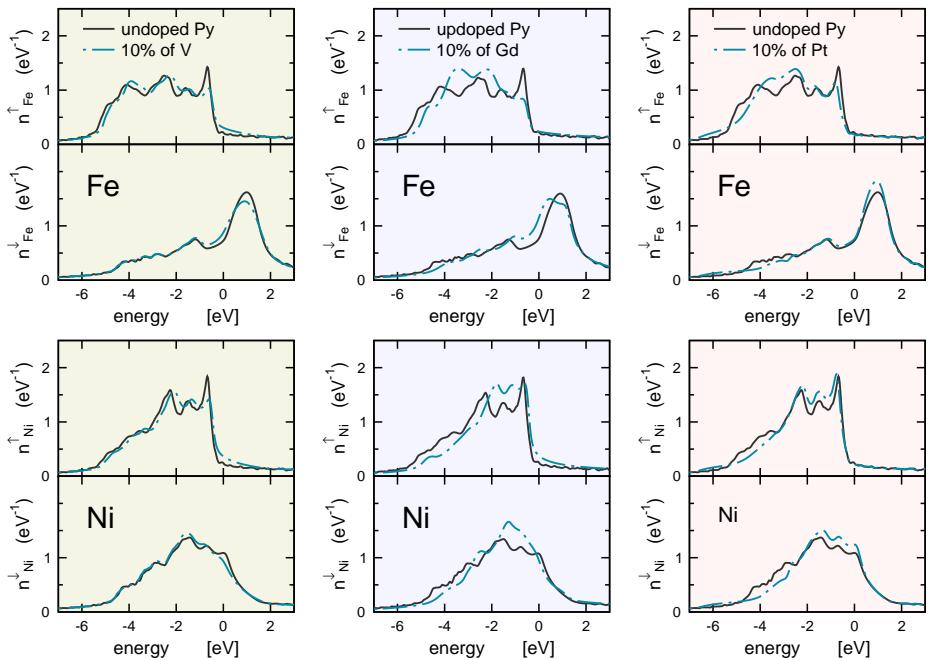}%
\caption{(Color online) Upper panels: Comparison of the DOS for Fe
  atoms in undoped Py and in Py doped by 10~\% of V (left), Gd (center),
  and Pt (right).  Lower panels: As in the upper panels but for Ni
  atoms.}
\label{fig-DOS}
\end{SCfigure*}

The next step is to inspect the electronic structure. The changes in
the DOS for the Fe and Ni atoms caused by the doping can be seen in
Fig.~\ref{fig-DOS}, where we show in the same graphs the data for
undoped Py and for Py doped by 10~\% of V, Gd, and Pt.  For the V dopant
the changes are minimal: the corresponding DOS curves are hardly
distinguishable from each other.  For the Pt dopant, the changes are
more significant.  Still larger but as a whole similar changes can be
observed also for the Gd dopant.  Based on Fig.~\ref{fig-DOS}, one
would infer that the hybridization between the electronic states of
the host and of the dopant is largest for the Gd dopant and smallest
for the V dopant.

\begin{figure*}[t]
\includegraphics[viewport=1.0cm 0.8cm 17.2cm 19.5cm]{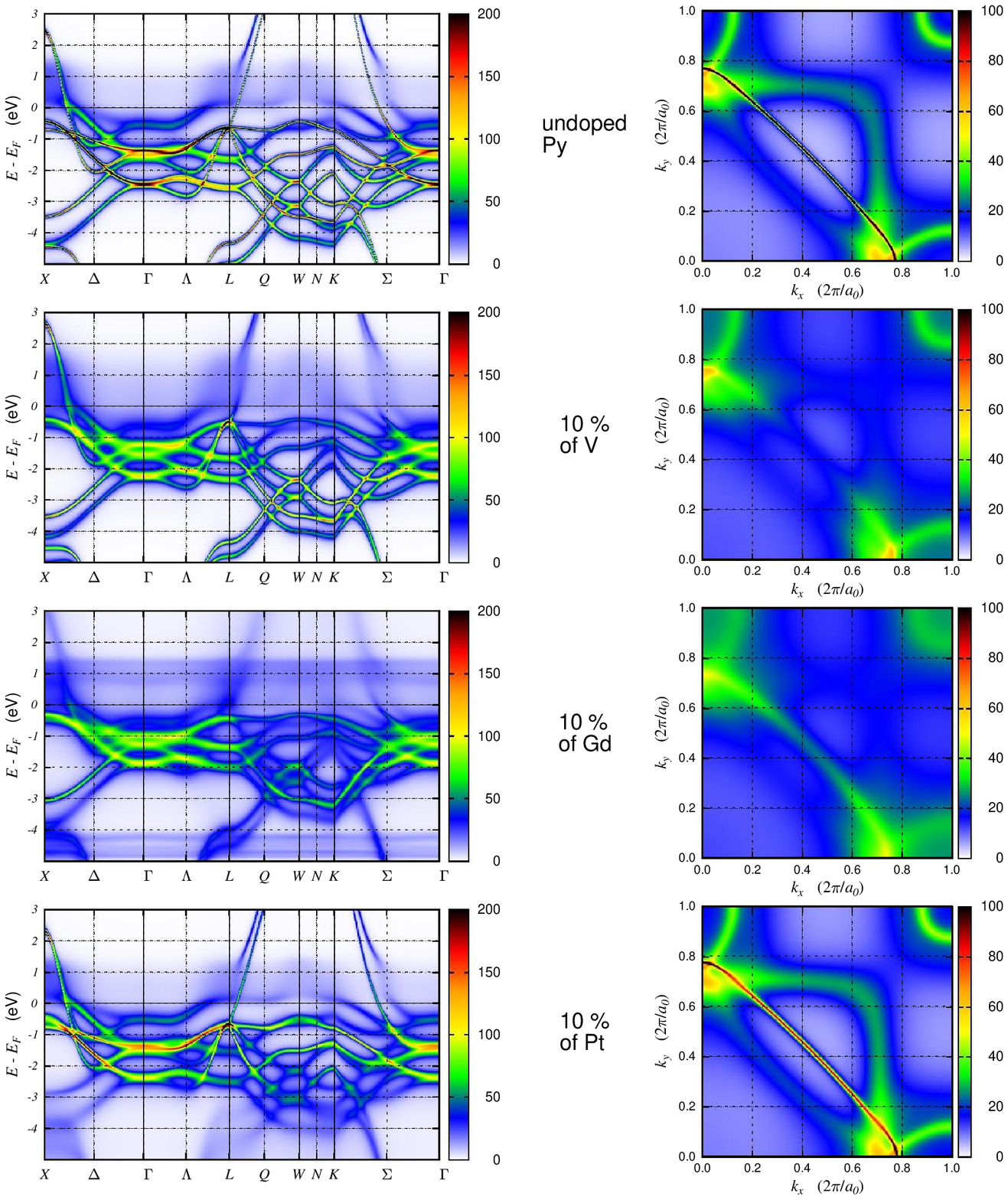}%
\caption{(Color online) Comparison of Bloch spectral functions for
  undoped Py and for Py doped by 10~\% of V, Gd, and Pt. }
\label{fig-bsf}
\end{figure*}

Reckoning the results discussed so far, there seems to be a difference
in the picture offered by inspecting the $J_{ij}$ constants in
Fig.~\ref{fig-jij} and by inspecting the DOS in
Fig.~\ref{fig-DOS}. The analysis of the coupling constants suggests
that the V and Gd dopants have a similar effect on the electronic
structure (which determines the $J_{ij}$'s) whereas the influence of
the Pt dopant is different. This follows also from spin wave stiffness
$D$ shown in Fig.~\ref{fig-theory-exper}.  On the other hand, the
analysis of the DOS suggests that qualitatively similar effects should
follow from introducing the Gd and Pt dopants whereas it is the V
dopant which differs in its effect on the electronic structure of
permalloy.

 A more thorough view on the electronic structure of alloys is
  provided by the Bloch spectral function $A(\bm{k},E)$, which can be
  viewed as a $\bm{k}$-resolved DOS. The effect of introducing 10~\%
  of V, Gd, and Pt into Py is shown in Fig.~\ref{fig-bsf}.  Left
  panels show $A(\bm{k},E)$ when the $\bm{k}$ vector is varied along
  straight lines connecting high-symmetry points in the first
  Brillouin zone, right panels show the Bloch spectral function at the
  Fermi level $A(\bm{k},E_{F})$ when the $\bm{k}$ vector spans a
  two-dimensional section of the $\bm{k}$-space (keeping $k_{z}$=0).
  By observing the $E-\bm{k}$ scans, one cannot say unambiguously
  which dopant introduces largest changes to the electronic structure
  of permalloy.  E.g., around the $\Gamma$ point the smallest changes
  to $A(\bm{k},E)$ are for the Pt dopant while changes introduced by V
  or Gd doping are larger.  On the other hand, if one focuses on the
  region around the $N$ point, doping by Gd has the largest impact
  whereas by V the smallest impact.  However, when discussing the spin
  wave stiffness, one has to keep on mind that it is associated with
  changes of electron energy caused by spin wave excitations
  characterized by small wave vector $\bm{q}$, and therefore
  contributed mainly by the electronic states close to the Fermi level
  $E_{F}$. As a result, the spin wave stiffness is sensitive first of
  all to the electronic structure around $E_{F}$.  The impact of
  doping on the states around $E_{F}$ is illustrated in the right
  panels of Fig.~\ref{fig-bsf}.  Here one can see that the changes
  introduced by doping by Pt are significantly smaller than changes
  introduced by doping by V or Gd.

\begin{table}
\caption{  Integrals of the differences of Bloch spectral
    functions at $E_{F}$ for doped and undoped systems, as defined in
    Eq.~\ref{eq-kint}. The units are arbitratry. }
\label{tab-kint}
\begin{ruledtabular}
\begin{tabular}{ldd}
 \multicolumn{1}{c}{dopant} &
 \multicolumn{1}{c}{concentration} &
 \multicolumn{1}{c}{concentration} \\
 \multicolumn{1}{c}{type} &
 \multicolumn{1}{c}{1\%} &
 \multicolumn{1}{c}{10\%} \\
\hline
V    &   4.49  &  20.18   \\ 
Gd   &   6.07  &  26.52   \\ 
Pt   &   1.38  &   7.62   
\end{tabular}
\end{ruledtabular}
\end{table}

 The influence of the doping on the electronic structure around
  $E_{F}$ can be quantified by integrating the difference of Bloch
  spectral functions for a doped system and for an undoped system.  We
  evaluated the integral 
\begin{equation}
  \int \! \dstd k \:
  \left|
  A_{\text{X}}(\bm{k},E_{F}) \, - \, A_{\text{undoped}}(\bm{k},E_{F})
  \right| \quad ,
  \label{eq-kint}
\end{equation}
 where $A_{\text{X}}(\bm{k},E_{F})$ and
  $A_{\text{undoped}}(\bm{k},E_{F})$ are the Bloch spectral functions
  for doped and undoped systems, respectively, and the integration is
  carried along the path
  $X$-$\Delta$-$\Gamma$-$\Lambda$-$L$-$Q$-$W$-$N$-$K$-$\Sigma$-$\Gamma$
  outlined at the horizontal axes of the left graphs in
  Fig.~\ref{fig-bsf}.  The results are shown in Tab.~\ref{tab-kint}
  for two dopant concentrations.  The largest changes in the
  electronic structure are introduced by Gd doping followed by still
  considerable changes introduced by V doping, while the changes
  introduced by Pt are small.

The picture emerging from analyzing Bloch spectra function is thus
different from the picture offered by the DOS.  The smallest changes
in $A(\bm{k},E)$ with respect to undoped Py are clearly for the Pt
dopant.  The changes introduced by V or Gd doping are bigger.  Based
on Fig.~\ref{fig-bsf} and Tab.~\ref{tab-kint} one would assess that
the changes in the electronic structure introduced by doping Py with
Pt are significantly smaller than the changes introduced by doping Py
with V or Gd.  Intuitively, this implies that for the Pt dopant the
changes in the values of $D$ will be smaller than for the V or Gd
dopants --- in agreement with the results presented in
Sec.~\ref{sec-the-exp}.  The Bloch spectral function thus offers a
more reliable picture than the DOS.


\section{Discussion}   \label{sec-diskuse}

Ab-initio calculations reproduce the experiment both as concerns the
dependence of $A_{\text{ex}}$ on the chemical type of the dopant and
as concerns the dependence of \aex\ on the dopant
concentration. Regarding the absolute numbers, there are some
differences, both between our theory and experiment and between our
theory and earlier calculations.  In particular our value of the spin
wave stiffness $D$ for undoped permalloy Fe$_{19}$Ni$_{81}$ of
576~meV~\AA$^{2}$ is higher than 515~meV~\AA$^{2}$ obtained by Yu
\ea\ \cite{YJK+08} via the tight-binding linear muffin-tin orbital
method or 522~meV~\AA$^{2}$ obtained by Pan \ea\ \cite{PCH+16} via the
KKR-Green's function method.  The reason for this is unclear, let us
just note that several convergence issues have to be addressed when
evaluating $D$; our settings are given in Sec.~\ref{sec-comput}.
Recent experimental values for $D$ of Py are 390~meV~\AA$^{2}$
\cite{Nak+83} and 440~meV~\AA$^{2}$ \cite{YAD+17}, i.e., less than the
theoretical values.  This is probably linked to problems with
describing the exchange coupling of Ni in terms of the coupling
constants $J_{ij}$ --- there is an even larger difference between
theory and experiment for $D$ of fcc Ni \cite{PKT+01,YJK+08}.
Besides, the difference between our values for \aex\ and the
experiment is probably affected also by the temperature-dependence of
$D$, which affects the interpretation of some experiments (as
mentioned in Sec.~\ref{sec-the-exp}).
 Another factor not accounted for by our calculations and
  possibly affecting the comparison between theory and experiment is
  the polycrystallinity of the experimental samples that were
  prepared by sputtering \cite{YPA+15,LCK+10,LCC+10,HGS+16}.  One
  should acknowledge that there may be problems also at the
  experimental side as the spread of values available in the
  literature is quite large.  A critical assessment would require a
  standalone study.  We just note for illustration that, e.g., in the
  case of Fe the available experimental data for $D$ include
  270~meV~\AA$^{2}$~\cite{Pau+82} as well as
  307~meV~\AA$^{2}$~\cite{LCL+84}, in the case of Ni
  398~meV~\AA$^{2}$~\cite{MP+85} as well as
  530~meV~\AA$^{2}$~\cite{Nak+83}, and in the case of Py
  335~meV~\AA$^{2}$~\cite{HHC+75} as well as
  390~meV~\AA$^{2}$~\cite{Nak+83}.  The differences between
  experimental values of \aex\ for Pt-doped permalloy reported by
  Hrabec \ea~\cite{HGS+16} and by Yin \ea~\cite{YPA+15}
  (cf.~Fig.~\ref{fig-theory-exper}) are thus not unusual.  Let us
  summarize that that discrepancies between theory and experiment are
  often observed for the spin wave and/or exchange stiffness and, so
  far, they are not fully understood.

The fact that the spin wave stiffness $D$ and the exchange stiffness
\aex\ are related through magnetization [see Eq.~(\ref{eq-A-from-D})]
which itself is affected by the doping means that there may be
situations where doping by two different materials will lead to a
similar $D$ but a different \aex\ (or vice versa).  In particular we
found that the spin wave stiffness $D$ for V-doped Py and Gd-doped Py
is very similar but there is a difference in the exchange stiffness
$A_{\text{ex}}$ which arises just from the difference in the
magnetization (cf.\ Fig.~\ref{fig-theory-exper}).

The influence of doping on the stiffness $D$ can be
 discussed just by considering 
the influence of the atoms in the first
coordination shell (Fig.~\ref{fig-D-J-fit}).  This might appear as
surprising reckoning the slow convergence of the sums in
Eqs.~(\ref{eq-simple}) or (\ref{eq-dsum}) \cite{PKT+01}. The
explanation might be that what we investigate in
Fig.~\ref{fig-D-J-fit} is just the {\em variation of $D$} with the
doping.  Eq.~(\ref{eq-fitJ1}), which we used for our analysis, was
derived by fitting the $D(J_{1})$ dependence relying on the values of
$D$ evaluated by extending the $\sum_{j}$-sum to the interatomic
distance as large as 20.5~$a_{0}$.  However, once the fit according to
Eq.~(\ref{eq-fitJ1}) has been established, one can predict how the
doping will influence the stiffness $D$ just by considering the
nearest-neighbors coupling.  Our fit has been verified for three quite
different dopants so presumably it can be used to assess the effect of
doping by other elements as well.  This can be helpful in current
efforts to manipulate spin-driven properties of materials.

Our results indicate that it does not really matter whether the
exchange coupling between the host atoms and the dopant atoms is
included in the stiffness calculation or not; the values of $D$ are by
about 5~\% smaller in the latter case but the trends are very similar
(see Tab.~\ref{tab-stiff-Gd} and Fig.~\ref{fig-howto}).  This
indicates that neither the polarizability of the dopant nor the
exchange coupling between the dopant atoms and the host atoms are
decisive factors for the spin wave stiffness $D$ (unlike what was
conjectured before, see for example Refs.~\cite{YPA+15,HGS+16}).  The
hybridization between the host states and the dopant states (or lack
of it) is more important than the dopant polarizability.
Interestingly, this hybridization has to be assessed not from the DOS
--- which contains only a limited information --- but from the Bloch
spectral function (cf.~Figs.~\ref{fig-DOS} and \ref{fig-bsf}).  The
fact that inspecting the Bloch spectra function provides a better
insight than inspecting the DOS alone can be seen as a demonstration
that the exchange coupling constants $J_{ij}$ reflect the full
electronic structure, including its $\bm{k}$-dependence.

Earlier studies indicated that the exchange coupling in 4$f$-electron
systems cannot be properly described within the local density
approximation or the GGA.  Employing the open core formalism was
suggested and tested for this purpose \cite{TKB+03}.  In our case we
found, nevertheless, that the stiffness of Gd-doped Py calculated
within the open-core formalism and within the GGA is very similar.
The reason for this might be that as the Gd atoms are here just
impurities, their $f$ electrons are localized ``naturally'', by the
lack of their hybridization with the host states. A further reduction
of the hybridization of the Gd $f$~states via the open core formalism
is thus not needed.

 Like most other calculations of the spin wave and exchange
  stiffness we rely on the ASA \cite{Kub+00,YPA+15,PCH+16,PKT+01}.
  For close-packed metallic systems such as those we are studying here
  this is not a serious limitation.  This was demonstrated, e.g., on a
  study of magnetic properties of disordered FePt where the CPA was
  applied both within the ASA and within a full-potential scheme
  \cite{KME+17}.


\section{Conclusions}   \label{sec-zaver}

Ab-initio calculations for doped permalloy indicate that the exchange
stiffness constant $A_{\text{ex}}$ decreases with increasing dopant
concentration.  This decrease is most rapid for the V dopant followed
by the Gd dopant, the slowest decrease is for the Pt dopant --- in
agreement with experiment.  The influence of the V-doping and the
Gd-doping on the spin wave stiffness $D$ is very similar, the
difference in the influence of the doping on the exchange stiffness
$A_{\text{ex}}$ comes from the differences in the magnetization
$M_{s}$ for V-doped and Gd-doped Py. The rate of change of the spin
wave stiffness upon introducing the dopants can be discussed 
by just considering the influence of the atoms in the first
coordination shell.  The hybridization between impurity and host
states is more important for the stiffness than the polarizability of
the impurity.


\begin{acknowledgments}
This work was supported by the GA~\v{C}R via the project 17-14840~S
and by the Ministry of Education, Youth and Sport (Czech Republic) via
the project CEDAMNF CZ.02.1.01/0.0/0.0/15\_003/0000358. Additionally,
financial support by the DFG via Grant No.~EB154/36-1 is gratefully
acknowledged.
\end{acknowledgments}


\appendix*

\section{Accuracy of \mm{\lim_{\eta\rightarrow 0}D(\eta)}\
  extrapolation} 
\label{sec-eta}

\begin{figure}
\includegraphics[viewport=0.5cm 0.7cm 9.0cm 5.7cm]{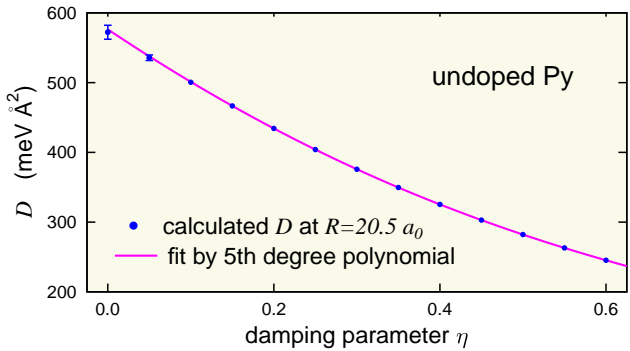}%
\caption{(Color online)
  Spin wave stiffness constant $D(\eta)$ for undoped
  Py evaluated by performing the summation in Eq.~(\ref{eq-dsum}) up
  to $R_{0j}^{\text{(max)}}=20.5a_{0}$ ($a_{0}$ is the lattice
  constant) for several values of the damping parameter $\eta$
  together with the fifth-degree polynomial used to extrapolate the
  data to zero damping. 
}
\label{fig-eta}
\end{figure}

The values of the spin wave stiffness $D$ presented in this
  paper rely on extrapolating data evaluated according to
  Eqs.~(\ref{eq-dalpha})--(\ref{eq-dsum}) for finite values of the
  damping parameter $\eta$ down to $\eta=0$.  This may be a tricky
  procedure. To present an analysis of all its aspects would be beyond
  the scope of this paper but the accuracy of our data can be
  illustrated on the case of undoped Py.  Fig.~\ref{fig-eta} displays
  original $D(\eta)$ data points together with the fifth-degree
  polynomial used for the \mm{\eta \rightarrow 0}\ extrapolation.  The
  $D$ constant obtained without any damping ($\eta=0$) is shown as
  well.  The errorbars depict errors caused by not yet fully damped
  $D(R)$ oscillations; they are discernible only for $\eta>0.1$ at
  this scale. The extrapolating polynomial was found by a
  least-squares fit within the interval \mm{\eta \in [0.15;1.00]}.
  One can see that the fit describes very well also the datapoints for
  $\eta=0.10$, $\eta=0.05$, and $\eta=0.00$ (within the numerical
  accuracy given by the $D(R)$ oscillations which are damped more and
  more as $\eta$ increases).  If the V, Gd, or Pt dopants are
  introduced, the procedure becomes even more robust: namely, further
  increase of the disorder increases the damping of the $D(R)$
  oscillations and the values of $D$ thus become numerically more
  stable.

 The extrapolation is not a unambiguous procedure.  One can
  always speculate about using a polynomial of a different degree for
  the fit or about varying a bit the interval of the $\eta$ values for
  which the polynomial is fitted.  Within these limits, the numerical
  accuracy of our values of $D$ is few units of meV~\AA$^{2}$.  This
  is sufficient for our purpose, especially given the fact that our
  focus is on the trends of the $D$ values evaluated always according
  to the same recipe.



\bibliography{liter_py-doped}

\end{document}